\documentclass[12pt]{iopart}

\expandafter\let\csname equation*\endcsname\relax
\expandafter\let\csname endequation*\endcsname\relax

\usepackage{amsmath}
\usepackage{amssymb}
\usepackage{graphicx}
\usepackage{bbm}
\usepackage{color}
\usepackage{subfigure}
\usepackage{epstopdf}

\def\mbf{\mathbf}

\def\i{\textrm{i}}

\newcommand{\one}{\leavevmode\hbox{\small1\kern-3.8pt\normalsize1}}

\begin{document}

\title{Driving Quantum Systems with Superoscillations   }

\author{Achim Kempf}
\address{Department of Applied Mathematics, University of Waterloo, 200 University
Avenue West, Waterloo, Ontario, N2L 3G1, Canada}
\ead{akempf@perimeterinstitute.ca}
\author{Angus Prain}
\address{Physics Department and STAR Research Cluster, Bishop's University, 2600 College St.,
Sherbrooke, Quebec, Canada J1M 1Z7}
\ead{angusprain@gmail.com}

\begin{abstract}
Superoscillations, i.e., the phenomenon that a bandlimited function can  temporary oscillate faster than its highest Fourier component, are being much discussed for their potential for  `superresolution' beyond the diffraction limit. Here, we consider systems that are driven with a time dependence that is off-resonance for the system, in the Fourier sense. 
We show that superoscillating sources can temporarily induce resonance during the period when the source is behaving superoscillatory. This observation poses the question as to how the system `undoes' the `false resonance' after the full source has acted and its band limitation is apparent. We discuss several examples of systems which might be capable of distilling the temporary excitation through some non-harmonic effects, such as dissipation or dispersion at high frequencies, opening up the possibility of low frequency detection of `fast' microphysics through superoscillations. We conclude that, either superoscillations really can beat the bandlimit and achieve superresolution (`kinematic superresolution') or the superoscillating high frequency is absorbed and we gain dynamical access to the physics of high frequency processes with low frequency signals (`dynamical superresolution').

%


\end{abstract}



\section{Introduction}

Surprisingly, there are functions which temporarily oscillate at frequencies that are much higher than their highest Fourier frequency component.  This phenomenon, known as superoscillation, was first discovered by Aharonov, Berry and others in the early 90s \cite{Berry:1990, berry1994faster, berry1994evanescent}. An analytic proof of their existence and a practical method for generating superoscillations was first given in \cite{Kempf:1999tq}. Superoscillations come with a tradeoff that explains why they have not been observed earlier: For a function to possess an interval with superoscillations, it must possess somewhere outside that interval slow oscillations that are of very large amplitudes. The required size of these amplitudes grows exponentially with the required length of the interval of superoscillations \cite{ferreira2006superoscillations,ferreira2002energy}. 

The occurrence of superoscillations in quantum wave functions, in quantum fields and in classical waves has been linked to a number of counter intuitive phenomena. For example, a particle described by a spatially superoscillating wave function can behave as if spring loaded: If only the  superoscillating part of the wave function passes through a slit, the particle will speed up upon passing the slit \cite{Kempf:2003vu}. Also, superoscillations in quantum fields have been proposed to be involved in the transplanckian problem of black hole physics \cite{reznik1997trans, rosu1996superoscillations}. Superoscillations in classical waves can be of potentially great practical significance. This is because the superoscillating parts of beams may be used to achieve superresolution, i.e., resolution beyond the diffraction limit, see e.g., \cite{zheludev}, with potential applications across the spectrum, for example in optical microscopy, microwave radar or terahertz imaging. 
In the present paper, we will argue that, beyond the purpose of superresolution, the use of waves that possess superoscillations can be of interest whenever a medium's transmission, absorption and reflection properties are significantly wavelength dependent. This is because the use of waves that possess superocillations then offers a win-win scenario. Namely, one desirable possibility is that (A) the material will treat the incident wave as composed of only long wavelengths (which it of course is by design) and reflect, absorb and transmit it accordingly. In this case, even the fast superoscillatory part of the signal will be reflected, absorbed and transmitted as if it were of long wave lengths. This then allows, for example, superresolution, as we will explain in more detail below. The other, also desirable, possibility is that (B) the material possesses fast internal dynamics that allows it to interact specifically with the fast superoscillating part of the incident wave. In this case, the material can reflect, absorb and transmit the fast superoscillating part of the incident wave in the way that it normally reflects, absorbs and transmits such short wavelengths. All cases are mixtures of the basic cases (A) and (B). 

For example, one may generate an optical beam composed of only red Fourier components which possesses blue superoscillating intervals. Assume that such a beam is directed at a surface that reflects red but absorbs blue. In case (A), all of the beam will be reflected. In case (B), the superoscillating part will be missing from the reflected signal. In case (A), we may use the reflected superoscillating part to determine the distance to the surface with an accuracy corresponding to the wavelength of the superoscillations, in a case of superresolution. In case (B), we have an opportunity to learn about some fast dynamics that happens in the material of the surface. For example, the material may be absorbent for blue because some molecular transition can absorb blue light. This excitation then decays because of intermolecular dynamics so that the energy eventually dissipates as heat through the material. In case (A), intuitively, a passing wave that possesses a superoscillatory stretch may temporarily excite the transition but will quickly de-excite it too so that all light is treated as if red. In case (B), intuitively, the superoscillatory stretch excites the transiton and before it can de-excite it again, inter-molecular dynamics will get hold of the excitation energy and start dissipating it. In this case, the blue superoscillating stretch will get absorbed. In case (B), we can therefore learn about some fast dynamics of the material. 

Our aim in this paper will be to study this interplay between cases (A) and (B), to show that the above intuition is correct. In particular, we will show that in case (A), the wave will indeed temporarily excite the medium to then quickly de-excite it. This then sets up the possibility of case (B), where this temporary excitation energy dissipates before the superoscillating wave can withdraw it again. Before, let us discuss more concrete examples. 

First, classical long wavelength waves that possess fast superoscillations may be useful in scenarios such as ground penetrating radar. There, a spatial resolution in the centimeter range (microwaves) is desired, for example, for the detection of landmines, while however such short wavelengths are very efficiently absorbed by humidity in the ground because of a resonance with rotational degrees of freedom of water molecules. Radar waves of significantly longer wave lengths are needed to achieve good ground penetration. In principle, we can choose such sufficiently long wavelength radar waves to be superoscillating in the centimeter range. In this case, if we are in scenario (A), these radar waves may still offer a sufficient spatial resolution. The question of whether this is an instance of case (A) or (B) can be tested experimentally and depends on the speed with which energy dissipates away from the rotational degrees of freedom of water molecules.  




A back of the envelope calculation suggests scenario (B) is physically reasonable in this case. The wavelength of microwaves which are absorbed by water is around 1cm to 30cm so let us choose $\lambda=10$cm microwaves for illustration. Then, assuming a superoscillating stretch of about 10 wavelengths, we have that the time period in which superoscillating microwaves are exciting the water molecules is
\begin{align}
\Delta t_\text{super}&=\frac{10 \text{cm}\times\lambda}{c}\\
&=\frac{10\times 10 \text{cm}}{1.3\times 3\times 10^8 \text{ms}^{-1}}\\
&= \frac{1}{4}\times 10^{-8}\text{s}\\
&\simeq 1\times 10^{-9} \text{s}
\end{align}
Water molecules are packed together such that the intra-molecular separation is approximately 1nm and the speed of sound in water is approximately $3\times 10^{3}$ ms$^{-1}$ so that the timescale of the interactions between water molecules is 
\begin{align}
\Delta t_\text{water}&=\frac{1\text{nm}}{10^{-3}\text{ms}^{-1}}\\
&=10^{-12}\,\text{s}. 
\end{align}
Therefore there is the opportunity for $\Delta t_\text{super}/\Delta t_\text{water}= 10^3$ water molecule interactions in the time it takes for the superoscillations to pass. This would seem ample time for the approximately harmonic rotational degrees of freedom of the water molecules to dissipate some of their resonant energy, spoiling the kinetic superresolution property.   This may open an  opportunity for sensitive measurements of the dissipation of energy in rotational degrees of freedom of water.



More generally, excitations near a local minimum of energy in \textit{any} physical system will behave quasi-harmonically, possessing a natural resonant frequency $\omega$ and ground state which can be excited by driving it locally near this resonance. Indeed, the optical properties of matter can be understood on these terms as being composed of a number of microscopic harmonic oscillators \cite{hopfield}. It is therefore of interest to study the basic interaction between superoscillations and harmonic and `almost-harmonic' oscillators. 


The structure of the paper is as follows. In Sec.~\ref{S:constructon} we outline a method for constructing superoscillating signals. In Sec.\ref{S:Harm} we will show that, as far as a linear harmonic oscillator is concerned, but also quite generally in perturbation theory for any quantum system with discrete energy levels, if a driving force does not contain the resonant harmonic frequency there is no energy imparted to the quantum system at late times. 


In Sec.~\ref{S:an_harm} we approach the problem of driving approximately harmonic systems with superoscillations, analysing three examples, the anharmonic oscillator, a dispersive oscillator and a parametrically driven oscillator as examples of non-harmonic behaviour. We outline some future outlook in Sec.~\ref{S:outlook}.

%
%
%
%

\section{Construction of superoscillations} \label{S:constructon}

In this section we review a versatile technique, which was first introduced in \cite{Kempf:1999tq}, to explicitly construct superoscillating functions, i.e., functions which oscillate in a short stretch much faster than their highest frequency component. 


We consider functions possessing compactly
supported Fourier transforms (band-limited functions), 
\begin{equation}
f(t)=\frac{1}{\sqrt{2\pi}}\int^\Omega_{-\Omega}d\omega\;
\tilde{f}(\omega)\text{e}^{\i\omega t}, \label{E:Fou}
\end{equation}
where $\tilde{f}$ is the Fourier transform of the function $f$ and $\Omega>0$ is the bandlimit. Let us ask whether we can constrain such a function to have the $N$
prescribed amplitudes,
\begin{equation}
f(t_i)=a_i, \quad \text{for} \quad i=1,\dots, N,
\end{equation}
with the idea to `force' the bandlimited function to follow the amplitudes of a high frequency waveform for some time. That is, from \eqref{E:Fou}, we ask of the function $f$ that
\begin{equation}
a_i=\frac{1}{\sqrt{2\pi}}\int^\Omega_{-\Omega}d\omega\;
\tilde{f}(\omega)\text{e}^{i\omega t_i}.
\end{equation}


To obtain a unique expression
for our superoscillating function we ask in addition for the function
to have minimal $L^2$ norm on the real line.  Writing this as an optimization
problem, we seek to minimize the functional $F[f]$ under the
constraints $G_i[f]=0$ where\footnote{Both $G_i$ and $F$ are
functionals of the infinitely many variables $\tilde{f}(\omega)$ and
$\tilde{f}^*(\omega)$ characterizing a given function $f$. Formally, the
variables $\tilde{f}(\omega)$ are independent of $\tilde{f}^*(\omega$) however
this is a redundancy since the two sets of Lagrange equations carry
the same information. We will thus only vary with respect to the
$f(\omega)$ variables. }

\begin{equation}
F[f]=\int d\omega\; \tilde{f}(\omega)\tilde{f}^*(\omega), \quad
G_i[f]=a_i-\frac{1}{\sqrt{2\pi}} \int
d\omega\;\tilde{f}(\omega)\text{e}^{\i\omega  t_i}.
\end{equation}
The standard Euler-Lagrange solution involves the Lagrange
multipliers $\mu_i$:
\begin{equation}
\frac{\delta F}{\delta \tilde{f}(\omega)}=\mu_i\frac{\delta
G_i}{\delta{\tilde{f}(\omega)}},
\end{equation}
where the sum over $i$ is implied. Specifically, this solution reads
\begin{equation}
\tilde{f}(\omega)=-\mu_i^*\frac{1}{\sqrt{2\pi}}\text{e}^{-\i \omega t_i}.
\end{equation}
Integrating both sides of this expression against the constraining
plane waves gives
\begin{align}
a_j&=-\mu_i^*\frac{1}{2\pi}\int^\Omega_{-\Omega}d\omega\; \text{e}^{\i(t_j-t_i)\omega} \\
&= -\mu_i^*\frac{1}{2\pi}\frac{\text{sin}\,\Omega(t_j-t_i)}
{\pi(t_j-t_i)}\\
&=:-\mu_i^* S_{ji},
\end{align}
where we have defined the matrix $S$ in the last line. Inverting $S$ we solve for the multipliers as
\begin{equation}
\mu_i^*=-S_{ji}^{-1}a_j.
\end{equation}
Therefore the unique solution is written as
\begin{align}
f(t)&=\frac{1}{\sqrt{2\pi}}\int^\Omega_{-\Omega} d\omega\;\tilde{f}(\omega)
\text{e}^{-\i\omega t}\\
&=\frac{S_{ji}^{-1}a_j}{2\pi}\int^\Omega_{-\Omega} d\omega\;
\text{e}^{\i\omega(t-t_i)}\\
&=S_{ji}^{-1}a_j
\frac{\text{sin}\;\Omega(t-t_i)}{\pi(t-t_i)}
\end{align}
where, again, we have employed a summation convention on repeated indices. The unique solution is seen to be a simple linear
combination of shifted sinc functions.  By choosing the amplitudes
$a_i$ judiciously to approximate any function we like we are able to
custom-make a superoscillating signal.  For example, given a shortest period of say $
T=4$ corresponding to a bandlimit of $\Omega=\pi/2$ one could prescribe the amplitudes $f(n)=(-1)^n$ for integer $n$, say between $-5$ and $5$, effectively
forcing the function to oscillate with period $T=2$ on this stretch, corresponding to a frequency of $\pi$, well outside the bandlimit. 

\begin{figure}
\centering
\includegraphics[scale=0.5]{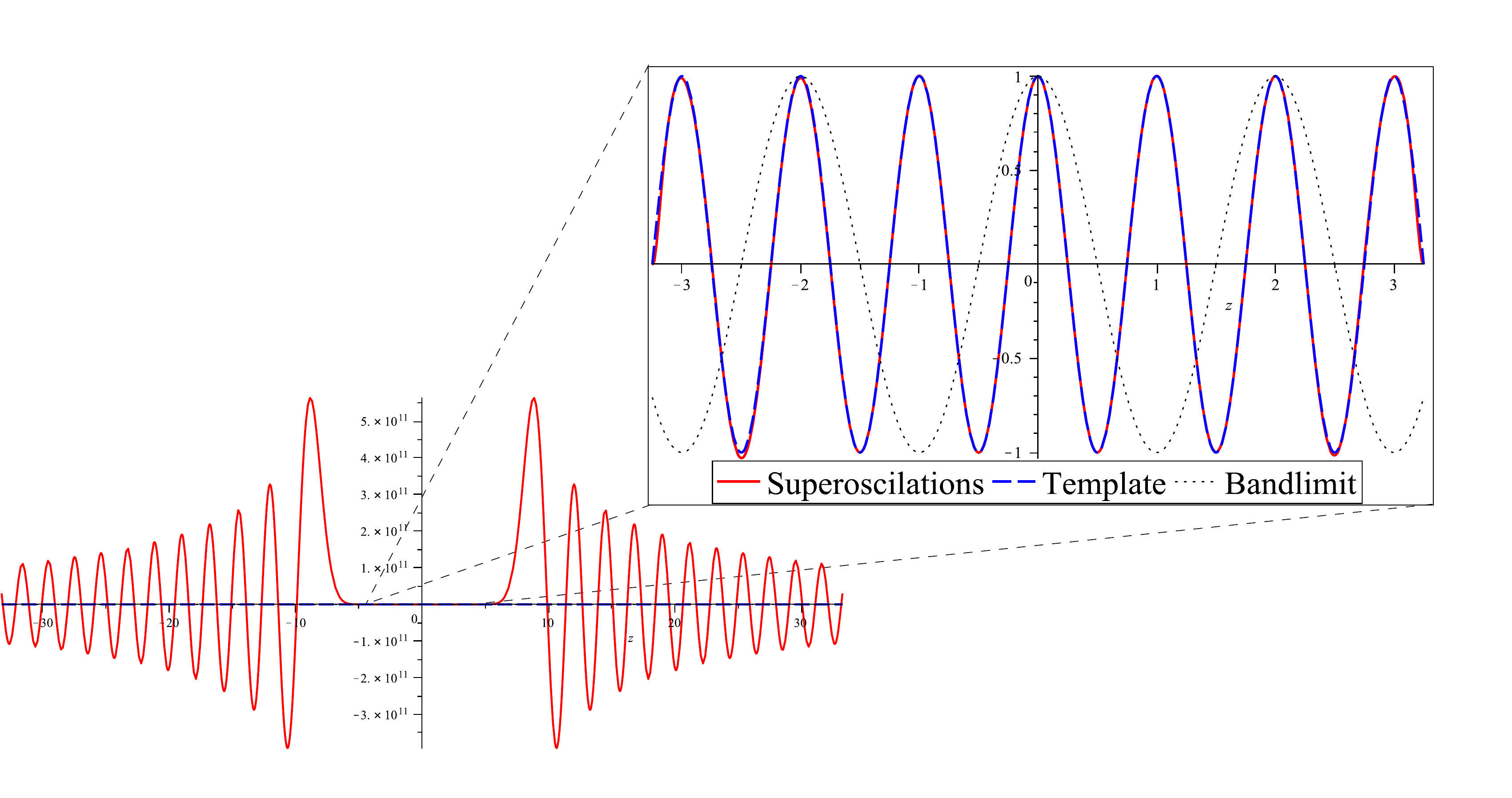}
\caption{The global picture of a superoscillating function, in red. The boxed sub-figure is windowed on the constraining interval where superoscillation occurs and we have also shown the shortest period Fourier component of the global signal.  \label{F:super_example_combined}}
\end{figure}

A shown in Fig.~\ref{F:super_example_combined} the price to pay for forcing a function to oscillate faster than its bandlimit is a large dynamic range of amplitudes between the constrained and unconstrained sections of the function. We see that, although we have constrained the function in Fig.~\ref{F:super_example_combined} to follow the plane wave cos $\pi t$ on the interval $[-4,4]$ the function possesses an amplitude of approximately $11$ orders of magnitude larger outside the constraining interval.

%
%
%

\section{Driving a quantum system with superoscillations \label{S:Harm}}

In this section we will investigate the problem of driving a system with a superoscillatory external agent. This will be done in two ways: Firstly we study the general problem of an n-level quantum system, working in perturbation theory; Secondly we provide a non-perturbative (exact) analysis of a driven harmonic oscillator. The results of both of these analyses corroborate the intuitive idea that an excitation will not remain in a quantum system at late times when the energy required to make the excitation is not present in the global frequency spectrum of the driving force. Our main result is to show that during resonant superoscillations the quantum system is temporarily excited, behaving as if truely driven at resonance. This we show in both the general n-level and harmonic oscillator cases and it is this behaviour that we have referred to as possibility (A) in the introduction. Possibility (B) would correspond to a physical system which behaves almost but not exactly as a harmonic oscillator, possessing the ability to dissipate or absorb some of the artificially induced excitations during superoscillations before they are de-excited by the driving force. 


\subsection{Exciting an n-level quantum system} \label{S:levels}

Consider a multi-level system described by the Hamiltonian $H_0(p,q)$ with discrete energy eigenstates $|n\rangle$ and eigenvalues $E_n$ such that $H_0|n\rangle =E_n|n\rangle$. Consider adding to such a system a time dependent interaction term $H_I(t)$ so that the full Hamiltonian is given by
\begin{align}
H(t)=H_0+H_I(t).
\end{align}
In the Schr\"odinger picture it can be shown that the system initially ($t=t_0$) in the state $|n\rangle$ evolves to the state at time $t$
\begin{equation}
|\psi (t)\rangle =\text{e}^{-\i H_0 t}\sum_m c_m(t)|m\rangle \label{E:coeffs},
\end{equation}
where the coefficients $c_m(t)$ satisfy the coupled differential equations 
\begin{equation}
\i\frac{d}{dt}c_m(t)=\sum_s V_{ms} \text{e}^{\i\omega_{ms}t}c_s(t) \label{E:c}
\end{equation}
where
\begin{equation}
V_{ms}=\langle m|H_I(t)|s \rangle, \quad \text{and}\quad \omega_{ms}=E_m-E_s. \label{E:actual_inner}
\end{equation}

For some special choices for the Hamiltonians $H_0$ and $H_I$ one is able to exactly solve the system \eqref{E:c} giving the probability of finding the system in state $|m\rangle$ at time $t$ as $|c_m(t)|^2$. An example of an exactly soluble system is the simple harmonic oscillator driven by a time dependent force $J(t)$ described by the interaction Hamiltonian $H_I=qJ(t)$.  We will discuss this solution in a following sub-section. 

In most cases, however, an exact solution to the interacting problem is not available even if the time-independent (non-interacting) problem can be solved exactly. In this case one can make progress by assuming that the interaction is proportional to a small parameter $\delta$ and a solution as a perturbative series in $\delta$ is easily obtained. It is a standard result that, if prepared in the initial state $|n\rangle$ at $t_0$, the probability amplitude to first order in perturbation theory of measuring the state to be $|m\rangle $ is given by 
\begin{align}
c_m(t)&=\delta_{mn}-\i\int^t_{t_0} dt'\,\text{e}^{\i\omega_{nm}t'}\langle m|H_I(t')|n\rangle\\
&=\delta_{mn}-\i \delta \langle m|q|n\rangle\int^t_{t_0} dt'\,\text{e}^{\i\omega_{nm}t'}J(t') \label{E:pert_result},
\end{align}
where in the last line we have specified the interaction Hamiltonian as the standard one, $H_I(t)=\delta q J(t)$. 

As the integral appearing in \eqref{E:pert_result} will be occur many times in what follows we make the following definition:
\begin{equation}
S_\omega(t):=\frac{-i}{\sqrt{2\omega}}\int^t_{-\infty}dt'\,J(t')\text{e}^{\i\omega t'},
\end{equation}
so that, in terms of $S_\omega$ we have
\begin{equation}
c_m(t)=\delta_{mn}+\delta \sqrt{2\omega}\,\langle m|q|n\rangle\, S_{\omega_{nm}}(t).
\end{equation}

\begin{figure}
\centering
\includegraphics[scale=0.5]{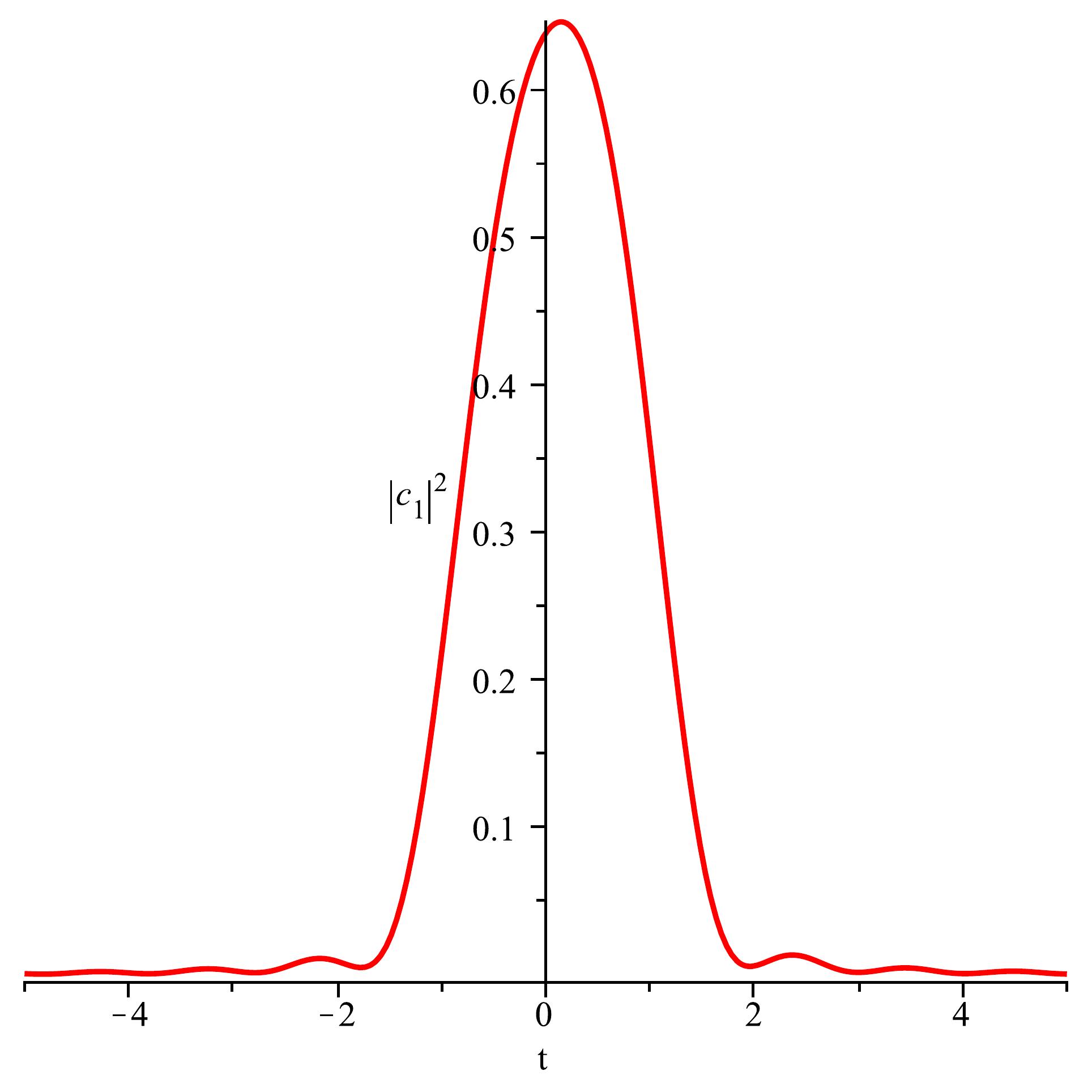}
\caption{The generic result of computing the Fourier transform up to time $t$ evaluated at $\omega$ of a superoscillating function bandlimited to contain only frequencies below $\omega/2$. This function represents various physical results in the text for the excitation of a quantum system possessing a resonance at frequency $\omega$ while being driven by an $\omega$-resonant superoscillating driving force bandlimited to $\omega/2$.  \label{F:transition}}
\end{figure}

From this result we see that if the driving force does not contain the energy difference $\omega_{mn}=E_m-E_n$ in its frequency spectrum then the system will be found to be in the state $|m\rangle$ at late times with zero probability since
\begin{equation}
c_m(t)\longrightarrow-\i \delta\langle m|q|n\rangle \,\tilde{J}(\omega_{nm})\quad \text{as} \quad t\rightarrow +\infty \label{E:pert_result2},
\end{equation}
where $\tilde{J}$ is the Fourier transform of $J$. It is then interesting to ask what the transition probability amplitude $c_m(t)$ looks like at intermediate times when the driving force is superoscillating at the frequency $\omega_{nm}$ but bandlmited below this frequency. The time-dependence of the coefficient $c_m(t)$ is all contained in the integral $S_{\omega_{nm}}(t)$, which is the `partial Fourier transform up to time $t$'. In Fig~\ref{F:transition} we plot the square modulus of this time dependence $|S_{\omega_{nm}}(t)|^2$ in the case where the driving force is bandlimited to frequencies $\omega<\omega_{nm}/2$ while superoscilating at $\omega_{nm}$ on an interval. 
We see that, indeed, the transition probability is essentially zero except in the superoscillating window where there is a sizeable non-zero probability to measure the system in an excited state.  

We interpret this result as follows: Globally the system returns to the initial ground state and no excitation occurs. Locally, during superoscillation, the system is excited and behaves as if driven at resonance by the resonant superoscillating driving force. This is the prototypical example of scenario (A) where only the low frequency physics of the bandlimited signal is represented globally and superresolution is possible. 

We point out however that it might be possible for a system to behave in a non-perturbative manner precisely during superoscillation in such a way that the temporary excitation is dissipated amongst some auxiliary coupled degrees of freedom, escaping this global conclusion. This we would refer to as scenario (B) and also leads to potentially desireable consequences.

\subsection{Forced harmonic oscillator}

The conclusion reached above is not exclusively confined to the realms of perturbation theory. Here we provide an exact analysis of a quantum system (the harmonic oscillator) described by scenario (A) exactly.   
The prototypical solvable quantum system is the forced harmonic oscillator.  The Hamiltonian is given by
\begin{equation}
H(p,q,t)=\frac{1}{2}p^2+\frac{1}{2}\omega^2q^2-J(t)q \label{E:exactlyH},
\end{equation}
where $J(t)$ is the time dependent driving force. 
Working in the Heisenberg picture we introduce the auxiliary operator $a(t)$ and its adjoint
$a^\dag(t)$ by
\begin{equation}
a(t)=\sqrt{\frac{\omega}{2}}\left(q(t)+\frac{i}{\omega}p(t)\right),
\quad a^\dag(t)= \left(a\right)^\dag(t) \label{E:change}
\end{equation}
in terms of which the Hamiltonian reads
\begin{equation}
H(t)=\omega\left(a^\dag(t)
a(t)+\frac{1}{2}\right)-\frac{1}{\sqrt{2\omega}}
\left(a^\dag(t)+a(t)\right)J(t). \label{Hamil}
\end{equation}
The equation of motion for $a$ is\footnote{It should be noted that, despite the fact that 
we merely take
complex linear combinations, the change of variables
$(q,p)\rightarrow(a,a^\dag)$ does not correspond to a canonical transformation - one that preserves
the canonical structure. That is, the Poisson bracket for these new
variables is $\{a,a^\dag\}=i$ in contrast to $\{p,q\}=1$. For this
reason the equations of motion are not the Hamiltonian equations
associated with the new variables i.e:
$\dot{a}\neq\partial_{a^\dag}H$ but instead are obtained directly
from those of $(p,q)$.}
\begin{equation}
i\dot{a}(t)=\omega a(t)-\frac{1}{\sqrt{2\omega}}J(t),
\end{equation}
with commutation $[a(t),a(t)^\dag]=\one$, solvable by an
integrating factor as
\begin{equation}
a(t)=\mbf{a}\,\text{e}^{-i\omega
t}+\one\frac{i}{\sqrt{2\omega}}\int^t_{t_0} \;
dt'\;J(t')\text{e}^{i\omega(t'-t)} \label{generator}
\end{equation}
just in case $[\mbf{a},\mbf{a}^\dag]=\one$.  Here, $\mbf{a}$ is a fixed operator understood as the initial
condition $a(t_0)$ and is the anihilation operator for the un-driven oscillator\footnote{There is an alternative way of obtaining
this solution if the use of an intergrating factor for operators
makes one a little uneasy: The most general family $\{a(t)\}_{t\in
\mathbb{R}}$ of operators satisfying the commutation $[a(t),
a^\dag(t)]=\one$ for all $t$ is parameterized by
$a(t)=\mbf{a}\,v(t)+\one u(t)$ where $[\mbf{a},\mbf{a}^\dag]=\one$.
Reinserting this into the equation of motion we obtain the relation
$\mbf{a}\left(i\dot{v}-\omega v\right)=\one\left(i\dot{u}-\omega
u+J/\sqrt{2\omega}\right)$ implying that both numerical factors
vanish.  An ordinary integrating factor may then be utilized for the
right hand factor to solve for $u$ which, combined with the requirement
of consistency with the commutation relations, gives the solution
\eqref{generator}.}.

From this exact solution we can compute the excitation of the oscillator at time $t$ as characterised by the number operator $N(t):=a^\dag(t) a(t)$ as
\begin{align}
\langle0|N(t)|0\rangle &= \frac{1}{2\omega} \left|\int^t dt'\, J(t')\,\text{e}^{\i\omega t'}\right|^2 \\
&=|S_\omega(t)|^2. \label{E:number}
\end{align}
This result is identical to the result plotted in Fig~\ref{F:transition} where it was shown to be proportional to the transition probability in a superoscillator driven n-level system. As anticipated, this result \eqref{E:number} corroborates our general perturbative treatment above but in an exact analysis. 

The complete solution is given by writing down an exact expression for the quantum state as a function of time. This is achieved by writing the initial ground state $|0\rangle$ in terms of the eigenstates $|n\rangle_t$ to the time dependent Hamiltonian $H(t)$ as \cite{Jacobson:2003vx}
\begin{equation}
|0\rangle = \text{exp}\,\left(-
\frac{1}{2}|S_\omega(t)|^2\right)\sum_{n=0}^\infty\frac{S^n_\omega(t)}{\sqrt{n!}}|n\rangle_t \label{E:jacob}
\end{equation}
revealing it to be a so-called `squeezed state'. This expression shows us that
\begin{equation}
{}_t\langle 0|0\rangle=\text{exp}\left(-\frac{1}{2}|S_\omega(t)|^2
\right) \label{result},
\end{equation}
implying the convergence of the states
$|0\rangle_t\rightarrow|0\rangle$ as $t\rightarrow\infty$
for any source whose Fourier decomposition does not contain the
resonant frequency $\omega$.

We conclude that the harmonic oscillator quantum system behaves as in scenario (A) since it is the function $S_\omega(t)$ which determines the response and $S_\omega(t)$ converges to the Fourier transform at late times. If the physics of reflection, absorption and transmission is exactly described by harmonic oscillators and one shines red light which is superoscillating in the blue part of the spectrum onto a material which reflects red but absorbs blue then the full red signal will be reflected including the superoscillating blue part. The interpretation is that the blue light is temporarily absorbed but is re-emitted through a subtle interaction with the non-blue part of the signal in such a way that the global frequency response is respected.

To draw a closer parallel with the general result in the previous section and the notation used there we can equivalently work in the Schr\"{o}dinger picture and compute the coefficients $c_n(t)$ in \eqref{E:coeffs} which represent the probability amplitudes for finding the system to be in the state $|n\rangle$ at time $t$ given the initial state $|0\rangle$. Using $q=(a^\dag+a)/\sqrt{2\omega}$ and \eqref{E:c} the $c_n$ satisfy
\begin{equation}
\i\frac{d c_n(t)}{dt}=J(t)\left[\text{e}^{\i\omega t}\sqrt{\frac{n}{2\omega}}c_{n-1}+\text{e}^{-\i\omega t}\sqrt{\frac{n+1}{2\omega}}c_{n+1}\right].
\end{equation}
This system is solved by the well known result \cite{feynman2012quantum}
\begin{equation}
c_n(t)=\text{exp}\left[\frac{-\i}{\sqrt{2\omega}}\int^t ds\,J(s)\text{e}^{-\i\omega s}\,S_\omega(s)\right] \frac{S_\omega^n}{\sqrt{n!}}
\end{equation}
and we have the exact solution for the wave function at all times
\begin{equation}
|\psi(t)\rangle = \sum_n c_n(t)\text{e}^{-\i E_n t} |n\rangle,
\end{equation}
where
\begin{equation}
E_n=\left(n+\frac{1}{2}\right)\omega.
\end{equation}

We see that the state $|\psi(t)\rangle$  in fact (weakly) converges to the initial ground state when the driving force does not contain the frequency $\omega$ since
\begin{align}
\langle 0|\psi(t)\rangle&= \text{exp}\left[\frac{-\i}{\sqrt{2\omega}}\int^t ds\,J(s)\text{e}^{-\i\omega s}\,S_\omega(s)\right]\\
& \longrightarrow \text{exp}\left(-|S_\omega|^2\right)\,\text{exp}\,\left(\i\phi(t)\right) \quad \text{as}\quad t\rightarrow +\infty,
\end{align}
where $\phi(t)$ is a real time dependent phase, implying convergence of the norm when $\tilde{J}(\omega)=0$ \footnote{The presence of the additional time dependent phase relative to the result \eqref{E:jacob} is related to the fact that in the Heisenberg picture we are free to choose a phase for the wavefunction at each time since this wave function is not required to satisfy any equation of motion. }.

\subsection{Discussion}
The specific question we have addressed in this section is:
{To what extent does a quantum system driven by a bandlimited signal that specifically does \emph{not}
contain the resonant frequency, but which nevertheless is constructed
to superoscillate at the resonant frequency, behave as if it were
truly driven by a resonant source?}

The question is non-trivial since the source is precisely constructed
to \textit{not} possess the resonant frequency and so should not excite
the system at all. This physical fact is displayed by the exact results above, generally in perturbation theory \eqref{E:pert_result} and specifically in exact harmonic oscillators \eqref{result}, that the system returns to the ground state asymptotically if the source does not contain
the resonant frequency and hence is not excited by the driving force. This result shows that the global form of the driving force is what matters in the long run in these cases. 

Interestingly, however, we also saw that during superoscillations the system becomes excited as if driven at resonance. This shows that the slowly varying external parts of the superoscillating signal are resonant with such a phase that they act just like the time-reverse of resonant amplification, namely by removing the induced excitations in order to preserve the global frequency response of the oscillator/quantum system.  This finely balanced scenario depends on a precise cancellation of effects which could be spoiled in a real quantum system. It is conceivable that fast microphysics processes might dissipate the induced excitation during superoscillations before it can be withdrawn again by the slowly varying external region.

\section{Epilogue: Beyond perturbation theory and harmonic oscillators \label{S:an_harm} }


In the previous section we saw two behaviors occurring for driven, exactly harmonic oscillators and generally in perturbation theory: (i) the system behaves globally with the appropriate frequency response -- if $\omega$ is not in the spectrum of $J$ then the system is not excited at late times; and (ii) temporarily, during resonant superoscillations, the system behaves locally as if driven at resonance.  

One can imagine at least two distinct ways in which new `internal'  high frequency degrees of freedom (e.g. non-linearities, anharmonicities etc.) can influence this behaviour. On the one hand we could have $f(+\infty)\neq 0$ even if $\tilde{J}(\omega)=0$ where $f$ is the classical position of the oscillator and $J$ is the driving force, the quantum analogue being that the Schr\"odinger state of the system does not converge at late times to the initial vacuum state.  It would then be interesting to compare the late time excitation $f(+\infty)$ (or, for example $\langle N(t) \rangle$ in the quantum case) between superoscillating and non-superoscillating driving, attributing any significant difference to a dissipation of the temporarily induced excitation into the new degrees of freedom.  On the other hand, one could attempt to directly excite the new (non-linear or internal) degrees of freedom themselves with superoscillations. This second possibility would require one to drive the approximately harmonic system with superoscillations at the new frequency scale, much faster than the natural resonance scale $\omega$. Below we will discuss both these possibilities.

In this section we will look at three generalisations to the harmonic oscillator system which possess new and additional frequency scales and how they might behave when driven by superoscillations. These are: i) a non-linear oscillator; ii) higher than second order differential equation; and (iii) a parametric driving or `dynamical Casimir' type scenario where the excitation spectrum itself is allowed to become time-dependent. Due to the intrinsic numerical difficulty of working with real superoscillations we leave their explicit solutions to a future study.

%

\subsection{Non-linear oscillator}

As we suggested in the introduction, all physical systems behave approximately harmonically near an energy minimum. What role do small non-linearities play in the conclusion that the harmonic oscillator returns to the ground state asymptotically? For example the effective potential for the physical pendulum can be approximated with a quartic potential
\begin{equation}
V(q)=m\omega^2(1-\cos(q))\simeq \frac{1}{2}m\omega^2 q^2-\frac{1}{24}m\omega^2 q^4 +\dots
\end{equation}
To be concrete we shall consider here the driven quartic modified anharmonic oscillator
\begin{equation}
H_\text{an}=\frac{1}{2}p^2+\frac{1}{2}\omega^2 q^2+  \lambda q^4-J(t)q,
\end{equation}
where $\lambda>0$ is a (not necessarily small) constant of dimension (mass)$^2$/(length $\times$ time)$^2$. Such a system and its equation of motion
\begin{equation}
\ddot{q}+\omega^2q +\lambda q^3=J
\end{equation}
has been extensively studied both perturbatively and non-perturbatively at the classical and quantum level \cite{PhysRevLett.77.4114,bender1969anharmonic,bender1973anharmonic} since the original work of Duffing in 1910. $H_\text{an}$ possesses a very rich phenomenology even at the classical level, including chaotic behaviour, requiring  subtle methods for its analysis. 

Even classically and for perturbatively small  $\lambda\ll1$ we can already see that the non-linear oscillator will not behave as in scenario (A). Here we will show that the ocsillator will not return to its initial state state of rest assymptotically even when the driving force does not contain the resonant frequency.  Assuming $J(t)\rightarrow 0$ as $t\rightarrow \pm\infty$ the perturbative ansatz $q(t)=q_0(t)+\lambda q_1(t)+\dots$ gives
\begin{align}
\ddot{q_0}+\omega^2 q_0&=J \\
\ddot{q_1}+\omega^2 q_1&=q_0^3.
\end{align}
Then, making use of the retarded Green function, we have the exact solutions
\begin{align}
q_0(t)&=\int^t_{-\infty} ds\, \frac{\sin\,\omega(t-s)}{\omega}J(s)\longrightarrow \sqrt{2\pi} \,\frac{\sin\,\omega t}{\omega}\,\tilde{J}(\omega)\quad \text{as}\quad t\rightarrow +\infty \label{E:1}
\end{align}
and
\begin{align}
q_1(t)&=\int^t_{-\infty}ds\,\frac{\sin\omega(t-s)}{\omega} q_0^3(s)\longrightarrow \i\sqrt{2\pi}\,\frac{\cos\,\omega t}{\omega}\,\widetilde{\left[q_0^3\right]}(\omega)\quad \text{as}\quad t\rightarrow +\infty.\label{E:2}
\end{align}
Here we have assumed for simplicity that $J(t)$ is an even function of $t$ and $\sim$ stands for the Fourier transform.  Crucially, \eqref{E:1} and \eqref{E:2} tell us that even if $J$ is $\Omega$-bandlimited such that $\Omega<\omega$ and hence $q_0(+\infty)=0$, the driving force for $q_1(t)$ is the $3\Omega$-bandlimited function $q_0^3$ (being the cube of the $\Omega$-bandlimited function $q_0$) so that, as long as $\Omega>\omega/3$, we can have $q_1(+\infty)\propto \widetilde{\left[q_0^3\right]}(\omega)\neq0$. It can be understood that the non-linearity breaks the superposition property of the harmonic oscillator, allowing for phenomena such as frequency mixing and the generation of new harmonic modes. 

One could interpret this classical result as indicating that the anharmonic oscillator has a new resonance channel at $\omega/3$ which is able to be excited by an $\Omega$-bandlimited driving force when $\omega/3<\Omega<\omega$, a departure from scenario (A) as discussed in the introduction. Such fractional resonances are a known feature of anharmonic oscillators beyond perturbation theory.  It would be interesting to drive such an oscillator with superoscillations at the new resonance scale $\omega/3$, observing the intermediate and late time excitation. 

At the quantum level the story is less clear and would require a deeper analysis. Below we outline a simple practical method which can be used to probe the quantum response non-perturbatively. 

\begin{figure}
\centering
\subfigure{
\includegraphics[scale=0.5]{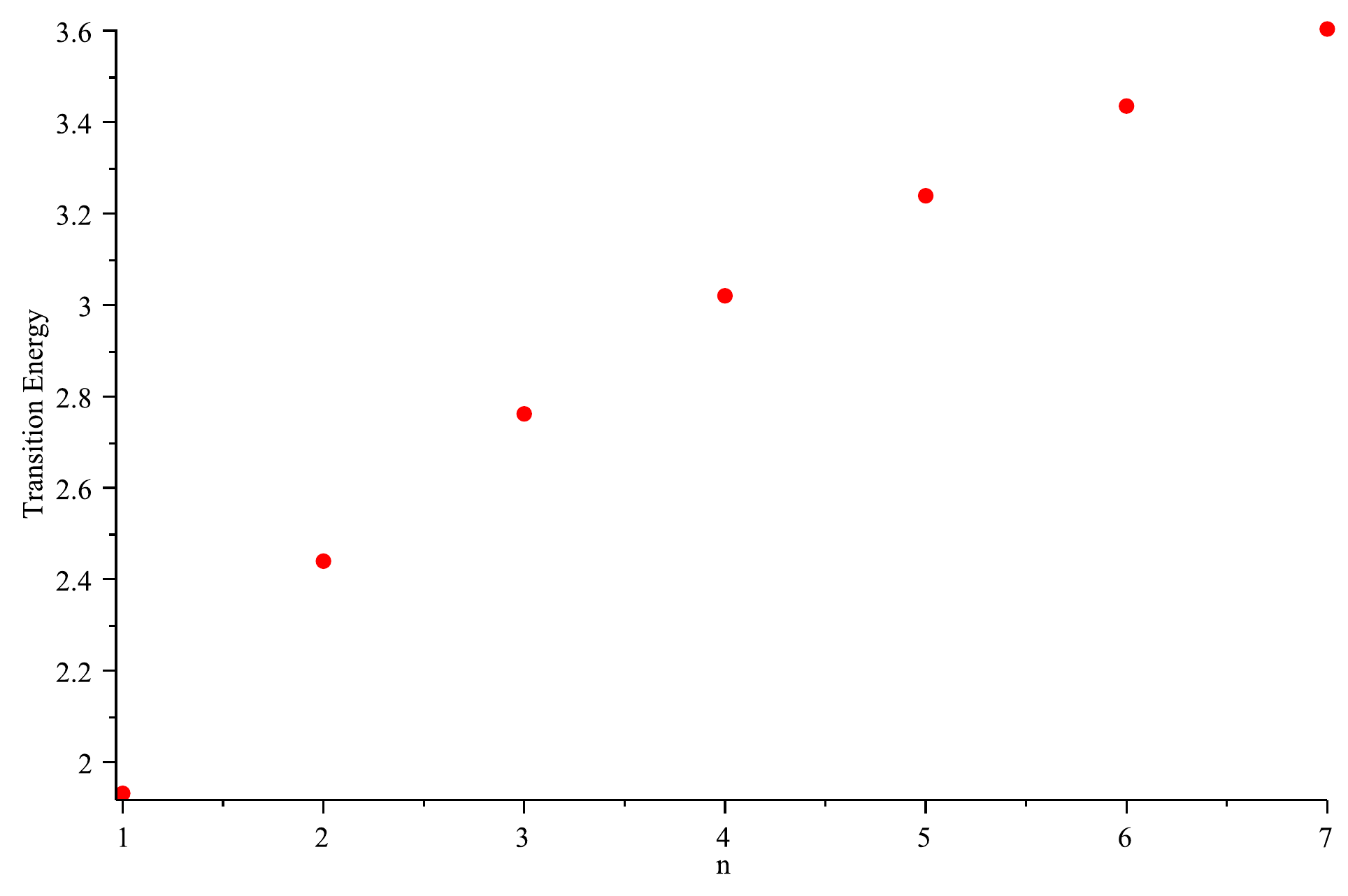}
}
\subfigure{
\includegraphics[scale=0.35]{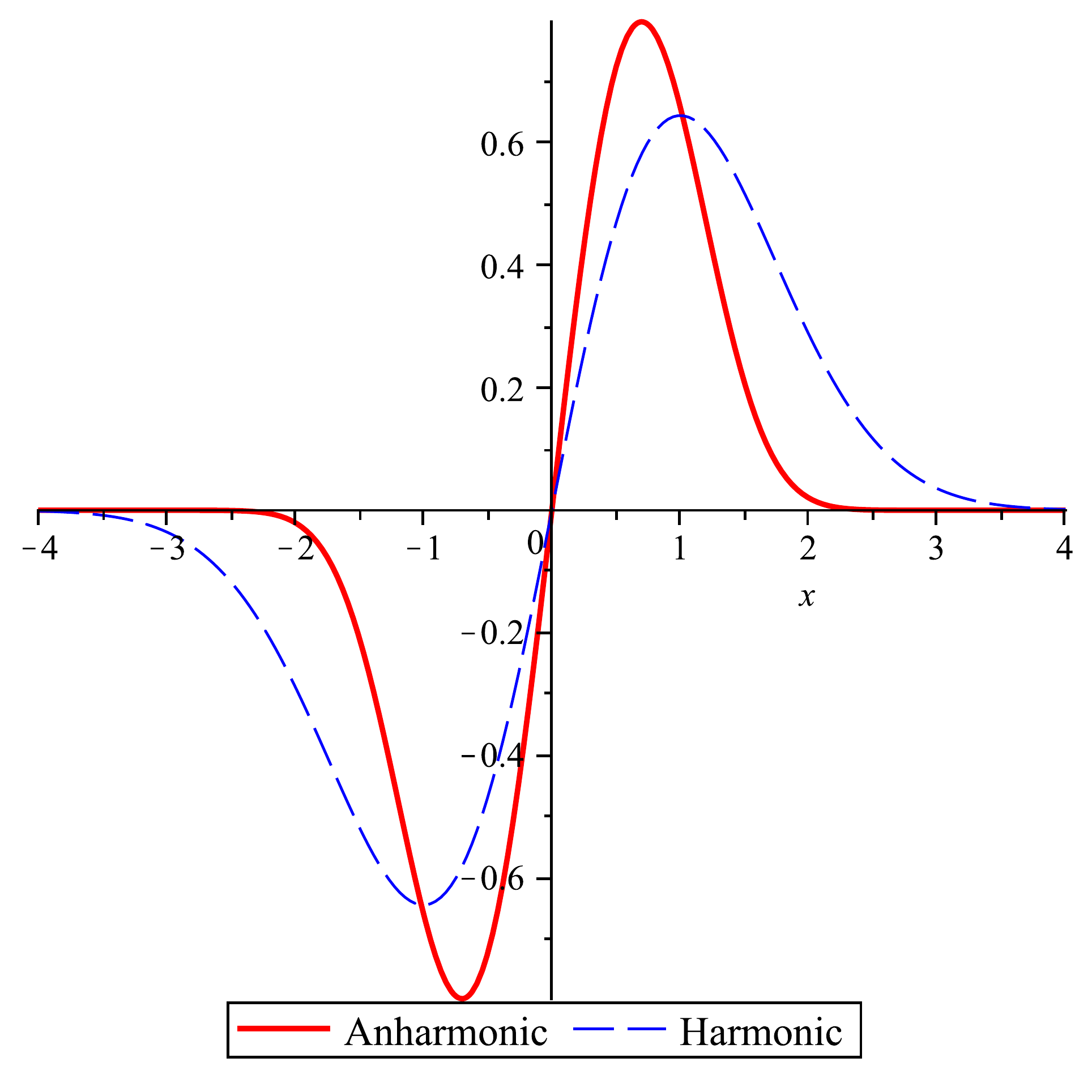}
\label{sub1}}
\subfigure{
\includegraphics[scale=0.35]{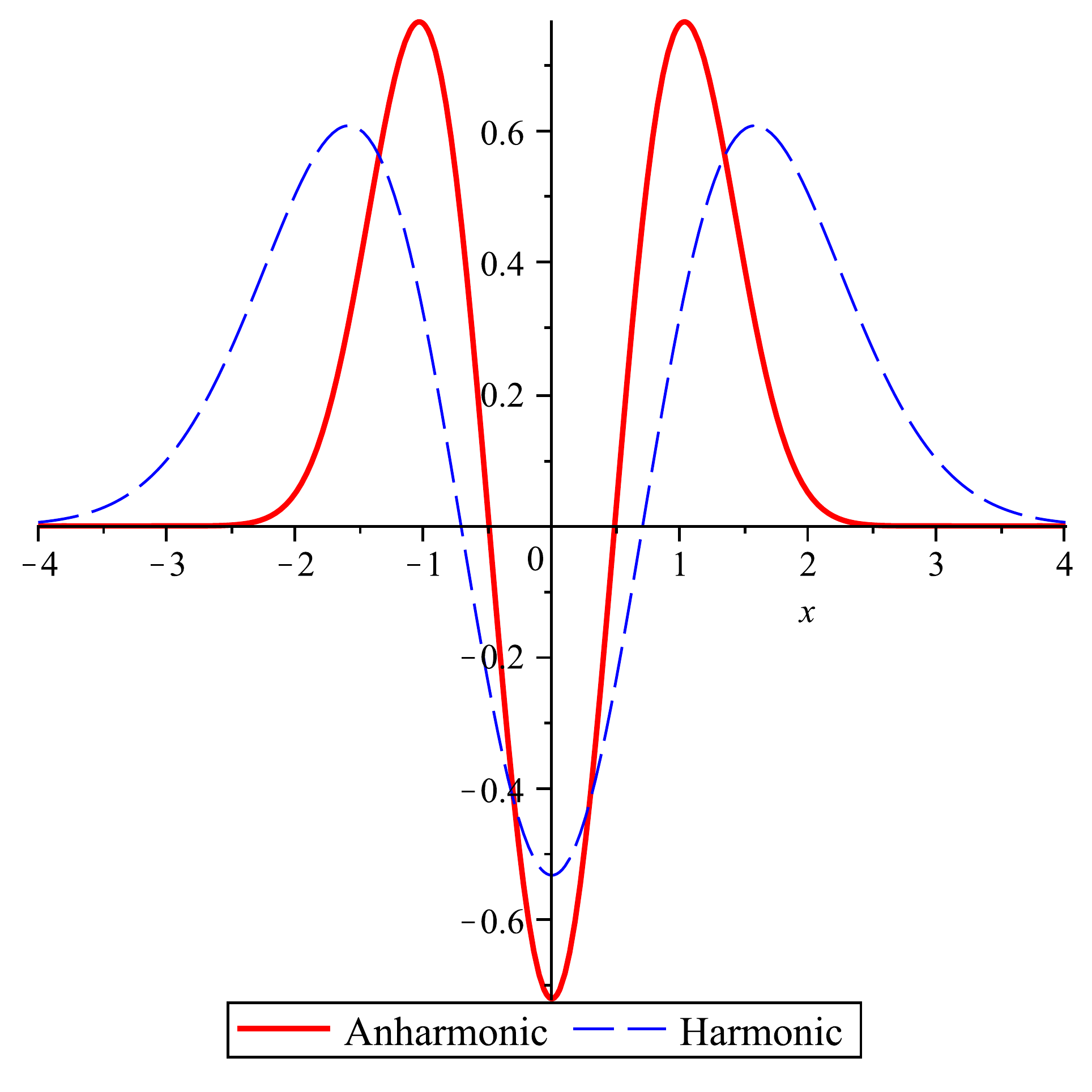}
\label{sub2}}
\caption{The transition energies $E_n-E_{n-1}$ as a function of $n$ for the anharmonic oscillator up to the 7th energy eigenvalue  and a comparison of the anharmonic energy eigenfunctions with the harmonic case. We chose $\lambda=1$ and $N=16$ to make these plots. \label{F:selection}}
\end{figure}

Consider the $N$ by $N$ matrix approximation  $(H^{N}_\text{an})_{nm}:=\langle n| H_\text{an}|m\rangle$ to the Hamiltonian $H_\text{an}$, where $|n\rangle$ are the (initial) eigenstates of the exactly harmonic Hamiltonian \eqref{E:exactlyH}. Diagonalising   $H^{N}_\text{an}$ we obtain an approximation to the first $N$ eigenstates $|n\rangle_\text{an}$ of $H_\text{an}$ and their energy eigenvalues. A selection of these approximate eigenstates and some energy eigenvalue gaps are shown in Fig.~\ref{F:selection} for the choice $\lambda=1$ and $N=16$.

The quantum dynamics is governed by the general set of coupled differential equations \eqref{E:c} involving the innerproducts ${}_\text{an}\langle m |q| n\rangle_\text{an}$ with respect to the exact anharmonic eigenstates $|n\rangle_\text{an}$. These can be computed  by inserting two (approximate) copies of the identity matrix:
\begin{align}
{}_\text{an}\langle n |q| m\rangle_\text{an}&= \sum_{r,s}^N{}_\text{an}\langle n| r\rangle\langle r|q|s\rangle \langle s| m\rangle_\text{an} \\
&=\sum_{r,s}^Na_{r,n}^*a_{s,m} \langle r |q|s\rangle \\
&=\sum_{r,s}^Na_{r,n}^*a_{s,m} \frac{1}{\sqrt{2}}\left(\sqrt{s+1}\,\delta_{r,s+1}+\sqrt{s}\,\delta_{r,s-1} \right) \\
&=\frac{1}{\sqrt{2}}\sum_s^N a_{s,m}\left(\sqrt{s+1}\,a^*_{s+1,n}+\sqrt{s}\,a^*_{s-1,n}\right),
\label{E:inner}
\end{align}
where $a_{ij}$ are easily obtained from the diagonalisation procedure of $H_{\text{an}}^N$ above. We are now in a position to solve the coupled equations \eqref{E:c}.

We would be interested in the case where $J(t)$ is superoscillating at the first transition $E_1-E_0$ while being bandlimited to, say, half this frequency. We should solve for the coefficient $c_1(t)$, observing its behaviour during superoscillations as well as asymptotically at $t=+\infty$. A non-zero $c_1(t)$ as $t\rightarrow +\infty$ would indicate a detection of superoscillations by the anharmonic oscillator, which should be expected based on our classical analysis above. In that case it would be interesting to compare the case of superoscillating and non-superoscillating $J$, observing if the asymptotic excitation is enhanced in the superoscillating case, due to a dissipation of the temporary excitation in the superoscillating stretch through other non-linear specific channels. Also of interest would be a comparison between the anharmonic and harmonic cases of the transition probability $|c_1(t)|^2$ at intermediate times during superoscillation which we expect also to be appreciably non-zero during superoscillations in the anharmonic case. 

One expects that for small $\lambda$ the $c_n(t)$ to be effectively coupled only to a small number of $c_m$ ($m\neq n$) in \eqref{E:c} as in the exactly harmonic case ($\lambda=0$) where there are only 2 non-zero contributions to the sum over eigenstates. Thus a good approximation can be expected by working in the truncated system for perturbatively small $\lambda$, projecting onto the first $N$ harmonic eigenfunctions. 

\subsection{Modified dispersion}

Another simple model which exhibits a high frequency new scale is a harmonic oscillator modified by a fourth order time derivative 
\begin{equation}
\left(\frac{1}{\Lambda^2}\frac{d^4}{dt^4}+\frac{d^2}{dt^2}+k^2\right)q=0,
\end{equation}
where $\Lambda>0$ is some high frequency new physics scale. When $k<\Lambda/2$ this model is equivalent to the so-called Pais-Uhlenbeck oscillator \cite{PhysRevLett.100.110402,PhysRevA.71.042110,Mostafazadeh:2010yw}
\begin{equation}
\frac{d^4 q}{dt^4}+\left(\omega_1^2+\omega_2^2\right)\frac{d^2 q}{dt^2}+\omega_1^2 \omega_2^2 q=0 \label{E:PU}
\end{equation}
and the system possess the two distinct roots $\omega_{1,2}$ to the dispersion relation
\begin{equation}
\frac{\omega^4}{\Lambda^2}-\omega^2+k^2=0. \label{E:disp}
\end{equation}
Real degrees of freedom are pairs $(\omega,k)$ which satisfy \eqref{E:disp}, corresponding to the intersection of a straight line and a lemniscate (sideways figure eight). They are interpreted as propagating dispersive waves with non-trivial group and phase velocities $v_g:=d\omega/dk$ and $v_p:=\omega/k$. 

The frequencies $\omega_{1,2}$ can be written in terms of the spring constant $k$ and new physics scale $\Lambda$ as 
\begin{align}
\omega_{1,2}&=\frac{\Lambda}{\sqrt{2}}\sqrt{1\pm \sqrt{1-\frac{4k^2}{\Lambda^2}}}\\
&\simeq \Lambda,\,\, k\quad \text{when} \quad\Lambda\gg k.
\end{align}
This dispersive model can be derived from the Lagrangian with a kinetic energy which also depends on the acceleration
\begin{equation}
L=-\frac{1}{2}\frac{1}{\Lambda^2}\ddot{q}^2+\frac{1}{2}\dot{q}^2-k^2q^2.
\end{equation}

If one were to drive such an oscillator with a superoscillating function which is superoscillating at the scale $\Lambda$ but band limited to frequencies well below $\Lambda$, what happens? Alternatively, if one drives the oscillator with superoscillations at the scale $k$, will any of the temporary excitations induced during superoscillations dissipate into the $\Lambda$ channel?

Classically we can show that such a system will not be excited asymptotically if the driving force is $\Omega$-bandlimited such that it does not contain either of the frequencies $\omega_1\sim k$ or $\omega_2\sim \Lambda$. One can see this in this case by simply taking the Fourier transform of the equation of motion for the driven system
\begin{equation}
\left(\omega^2-\omega_1^2\right)\left(\omega^2-\omega_2^2\right)\tilde{q}(\omega)=\tilde{J}(\omega)
\end{equation}
and hence 
\begin{equation}
q(t)=\int^\Omega_{-\Omega}\frac{d\omega}{\sqrt{2\pi}}\,\frac{\tilde{J}(\omega)}{(\omega-\omega_1)(\omega-\omega_2)}\,\text{e}^{\i\omega t}\longrightarrow 0 \quad \text{as}\quad t\longrightarrow +\infty, \label{E:limit}.
\end{equation}
The limit vanishes since the factor multiplying the exponential in the integrand is completely regular on $[-\Omega,\Omega]$ when $\Omega<\omega_{1,2}$. The Greens function in the intergrand can be re-expressed as
\begin{equation}
G(\omega)=\frac{1}{\omega_1^2-\omega_2^2}\left[\frac{1}{\omega^2-\omega_1^2}-\frac{1}{\omega^2-\omega_2^2}\right],
\end{equation}
showing that this system is equivalent to two independent non-interacting exactly harmonic oscillators (with resonant frequencies $\omega_1$ and $\omega_2$ respectively). Therefore no dissipation can occur from the $k$ degree of freedom to the high frequency one at $\Lambda$ unless we modify the system to include a coupling between these two degrees of freedom. This possibility is considered in the article \cite{pavsic}. 

The result \eqref{E:limit} is also an intriguing result in itself as it implies that a bandlimited function which does not contain either of $\omega_1$ or $\omega_2$ but which is superoscillating at the new physics scale $\omega_1\simeq \Lambda$ will excite the oscillator temporarily during superoscillations with high probability (recall the result \eqref{E:number}) while leaving it assymptotically in the future un-excited.  Superoscillating functions therefore appear able to detect microphysics (in this case a high frequency new scale due to modified dispersion) by monitoring  the intermediate-time excitation probability. This would be a modified version of scenario (A) with the additional feature of intermediate stage (B) type-behaviour. Certainly, driving a system well above its fundamental resonance $\omega$ (with superoscillations) and observing a non-trivial excitation would be surprising indeed!

\subsection{Parametric excitation}

An oscillator can also be excited by a time variation of the spring constant described by the equation of motion
\begin{equation}
\left[\frac{d^2}{dt}+\omega(t)^2\right]q=0.\label{E:para}
\end{equation}
Here the energy spectrum itself becomes time dependent and therefore we are outside the regime in which the results of Sec.~\ref{S:levels} apply. 


Like the case $\omega(t)=$ constant, the Hamiltonian is diagonalised at each time by instantaneous energy eigenstates 
\begin{equation}
|n\rangle_t:=\frac{1}{n!}a^\dag(t)^n|0\rangle_t
\end{equation}
where
\begin{equation}
a(t):=\sqrt{\frac{\omega(t)}{2}}\left(q+\frac{\i}{\omega(t)}p\right).
\end{equation}
Since the spectrum is time dependent, the quantum states $|0\rangle_t$ are all different.  Avoiding the issue of defining particles in a time dependent setting, one can choose $\omega(\pm\infty)=$constant, denoting the $t\rightarrow \pm\infty$ vacuum states as $|0\rangle_\text{in,out}$ respectively. It is a standard result \cite{birrell1984quantum} that
\begin{equation}
{}_\text{in}\langle 0|0\rangle_\text{out}\neq 0 \label{E:inout}
\end{equation}
whenever $\omega(t)$ is not identically constant even when $\omega(-\infty)= \omega(+\infty)$ \textit{independently of how $\omega(t)$ varies in time}. 

The non-zero value of this inner-product is traditionally denoted by $|\beta|^2$, is given by 
\begin{equation}
{}_\text{in}\langle 0|0\rangle_\text{out}=|\beta|^2,
\end{equation}
and is known as one of the two Bogoliubov coefficients (the other being denoted $\alpha$). The Bogoliubov coefficients relate the bonafide annihilation and creation operators which exist at $-\infty$ (labelled by $a,a^\dag$) and $+\infty$  (labelled by $b,b^\dag$) where $\omega(t)$ is constant\footnote{Compare this result to \eqref{generator} where the operator $a(t)$ acts appropriately as an annihilation operator in \textit{both} asymptotic regions when $J$ does not contain the frequency $\omega$.}. The relationship is given by the linear transformation 
\begin{equation}
b=\alpha a-\beta^*a^\dag.
\end{equation}
Explicitly $\beta$ is found using the exact solution $q_0(t)$ to the c-number version of \eqref{E:para} possessing the initial condition
\begin{equation}
q_0(t)\longrightarrow \frac{\text{e}^{-\i\omega t}}{\sqrt{2\omega}} \quad \text{as}
\quad t\rightarrow -\infty,
\end{equation}
propagating this exact solution through the non-trivial time-dependent region and decomposing it in the future static region $t=+\infty$ into plane waves 
\begin{equation}
q_0(t)\longrightarrow \alpha\,\frac{\text{e}^{-\i\omega t}}{\sqrt{2\omega}}+\beta\,\frac{\text{e}^{\i\omega t}}{\sqrt{2\omega}} \quad \text{as}\quad t\rightarrow +\infty. 
\end{equation}

In this parametric driving case the assymptotic ($t=+\infty$) excitation of the oscillator is characterised by the number operator associated with the Bogoliubov transformed creation operators $N_b=b^\dag b$ and is simply given by the coefficient $|\beta|^2$.
%
%
The important difference with the simple driven oscillator is that here the Fourier spectrum of $J$ does not determine the asymptotic excitation of the solution.  Nevertheless, it can be shown that if $\omega(t)$ is periodic  about a constant background value (say $\omega_0$) with frequency approximately twice this background value (that is, at $2\omega_0$) then the oscillator experiences an exponential increase of its energy, known as a parametric resonance. A similar but less severe resonance occurs if $\omega$ has period $\omega_0$ itself, but this resonance is less surprising.

It would be very interesting to check how the parametrically driven oscillator behaves when driven with superoscillations at the frequency $2\omega_0$ given that outside a small band around this parametric resonance, no exponential increase in energy is expected.

\section{Prospects and outlook \label{S:outlook}}

In this article we considered in perturbation theory the scenario of a quantum system that starts in the ground state and is then driven off-resonance (i.e., without the resonance frequency being present in the driving force's Fourier spectrum) but with the driving force exhibiting temporarily  superoscillations at the resonance frequency. We found that a system driven in this way i) locally, namely during superoscillations, behaves genuinely as if driven at resonance and ii) behaves globally with respect to the full Fourier spectrum of the driving force, as if driven off-resonance, namely it returns to the ground state at late times.  This perturbative result is exhibited exactly by the driven quantum harmonic oscillator. The desirable physical consequence of this behaviour is spatial  superresolution (case (A)). The understanding is that hidden inside the slowly varying lobe sections of the superoscillating driving force are resonant frequencies which first provide and consequently remove the energy which is removed and then added to the oscillator during superoscillations. These observations formed the main results of this work.  

However the results posed further questions and we were led to ask whether other behaviours than case (A) are possible.  The case-(A) involved fine-tuned cancellations which seem to sensitively rely on the assumption  that none of the temporary excitations during superoscillations are distilled or dissipated into other degrees of freedom before the remainder of the signal withdraws that energy again from the system. The slowly varying external lobes of a superoscillating function act to precisely de-excite the induced resonant excitations of the system, opening the possibility that, had we removed the superoscillation induced excitations before the lobes act, further excitations of the system might be possible, induced by the lobes which appear also to contain the appropriate high frequency. This could be interpreted as a ``double harvesting" of energy at the resonance scale -- an energy scale which is not present in the spectrum of the driving force! In this way superoscillations would open a window to probing high frequency internal dynamics using low frequency observation, which we have called dynamical superresolution (case (B)).  

One may also look at mixtures of cases (A) and (B). Imagine we are interested in a quantum system which we think has only low energy degrees of freedom and that we drive it with superoscillations way above any known resonance. If the system possesses a new energy scale at the superoscillation frequency we will notice the system becoming excited temporarily during this stretch, indicating an energy transition at that frequency. Consequently the system will de-excite as the final lobe section of superoscillations acts (assuming the perturbative result \eqref{E:pert_result} is appropriate) but we will have learned about some high frequency dynamics using only low frequency driving forces. 

We have discussed modifications to the basic harmonic oscillator under superoscillatory driving in an attempt to model case (B) or mixtures of case (A) and case (B). We plan on making a more complete dynamical analysis of these modified oscillator systems in the future.



It should be very interesting to extend the present results in several direction. For example, one may carry out further numerical investigations of equations which describe the various modifications, (perturbative or not at the level of the free Hamiltonian) to the basic systems we have studied here. Our general result in time dependent perturbation theory can be circumvented in a non-perturbative (in the driving force) treatment. Also, it is tempting to explore various experimental setups in various parts of the electromagnetic spectrum.  One example would be the production of microwave signals that are bandlimited so that they can pass through water but that possess a superoscillating stretch whose frequency is high enough to be efficiently absorbed by water. One could then probe to what extent behaviors of type A or B occur. Such an experiment would involve the challenge to generate superoscillating electromagnetic waves in the microwave part of the spectrum.  
\bigskip\newline
\bf Acknowledgements: AK acknowledges support from the NSERC Discovery Program. \rm

\vspace{5mm}

\bibliographystyle{utphys}
\bibliography{superbib}

\end{document}